\documentclass[prl,twocolumn,superscriptaddress,nofootinbib,longbibliography]{revtex4-2}
\usepackage{amsmath,amssymb}
\usepackage{graphicx}
\usepackage{xcolor}
\usepackage[colorlinks=true,linkcolor=blue,citecolor=blue,urlcolor=blue]{hyperref}
\usepackage{orcidlink}
\usepackage{cleveref}   
\usepackage[normalem]{ulem}

\newcommand{\figfile}[2][\columnwidth]{%
  \IfFileExists{#2}%
    {\includegraphics[width=#1]{#2}}%
    {\fbox{\parbox[c][0.6\columnwidth][c]{0.92\columnwidth}{\centering
       \textbf{[figure not in this checkout]}\\[2ex]
       \texttt{\footnotesize\detokenize{#2}}}}}%
}
\newcommand{\dd}{\mathrm{d}}
\newcommand{\cL}{\mathcal{L}}
\definecolor{darkpurple}{RGB}{85,26,139}

\begin{document}

\title{LZ Nuclear-Recoil Excess from Boosted Light Magnetic Dipole-dipole Dark Matter}

\author{Jin-Han Liang\orcidlink{0000-0002-6141-216X}}
\email{jinhanliang@wyu.edu.cn}
\affiliation{School of Applied Physics and Materials, Wuyi University, Jiangmen 529020, China}

\author{Zuowei Liu\orcidlink{0000-0002-7662-9206}}%https://orcid.org/0000-0002-7662-9206
\email{zuoweiliu@nju.edu.cn}
\affiliation{Department of Physics, Nanjing University, Nanjing 210093, China}

\author{Van Que Tran\orcidlink{0000-0003-4643-4050}}
\email{vqtran@phys.ncts.ntu.edu.tw} 
\affiliation{Physics Division, National Center for Theoretical Sciences, National Taiwan University, Taipei 106319, Taiwan}
\affiliation{Phenikaa Institute for Advanced Study, Phenikaa University, Nguyen Trac, Duong Noi, Hanoi 100000, Vietnam}

\author{Yongheng Xu\orcidlink{0000-0001-7094-5534}}
\email{yongheng.xu@fys.uio.no}
\affiliation{Department of Physics, University of Oslo, Box 1048, N-0316 Oslo, Norway}
\affiliation{Department of Physics and Astronomy, University of California, Los Angeles, CA 90095-1547, USA}

\begin{abstract}

The LZ collaboration has reported a  
nuclear-recoil excess near 248 keV 
with a global significance of $2.6\sigma$. 
Although halo dark matter with a magnetic dipole-dipole interaction and a TeV-scale mass 
provides the best fit to the excess among 
the interactions considered by LZ, 
it predicts a considerable number of events at lower recoil energies, 
where no excess is observed. 
We show that a boosted velocity distribution can alleviate this tension and provide a better fit to the LZ recoil spectrum. Moreover, the boost opens up the possibility of explaining the excess with much lighter dark matter, with masses down to the GeV scale. 
We demonstrate these features first in a model-independent analysis and
then realize them in a concrete dark matter model, in which halo
dark matter annihilates into on-shell mediators that subsequently 
decay into boosted dark-sector particles. 
Our results demonstrate that boosted dark sector particles 
provide a viable interpretation of the LZ excess.

\end{abstract}

\maketitle

%%%%%%%%%%%%%%%%%%%%%%%%%%%%
\section{Introduction}
%%%%%%%%%%%%%%%%%%%%%%%%%%%%
Direct searches for dark matter (DM) have traditionally focused on
nuclear recoils below $\sim100$~keV, as expected for non-relativistic
Galactic DM with characteristic velocities $v\sim10^{-3}c$.
The LUX-ZEPLIN (LZ) collaboration has recently extended its search to
recoil energies up to $E_R\simeq270$~keV with an exposure of
2.84~tonne-years~\cite{LZ:2026axp,LZtalk2026}.
A single event was observed at a nuclear recoil energy 
$E_R=248\pm23\, (\mathrm{stat})\pm23\, (\mathrm{sys})$~keV, 
against an expected background of only $0.0106\pm0.0008$ events, 
corresponding to a local (global) significance of 
$3.4\sigma$ ($2.6\sigma$). 
This event therefore provides an intriguing hint of a possible DM signal. 
However, the unusually high recoil energy and the absence of a corresponding
low-energy excess pose a distinctive challenge for a DM interpretation: 
a viable signal should produce an ${\cal O}(1)$ event near
$248$~keV without simultaneously generating a sizable population of
lower-energy recoils.

LZ identified two particularly favorable DM interpretations of this
spectral feature~\cite{LZ:2026axp}: endothermic inelastic scattering
with a DM mass above several hundred GeV and a mass splitting of order
a few hundred keV, and elastic magnetic dipole-dipole (MDD) scattering
with a best-fit mass near $1$~TeV.
The former possibility has motivated extensive studies of
quasi-degenerate Higgsinos with $m_\chi\simeq1.1$~TeV and
$\delta\sim300$--$400$~keV~\cite{Fan:2026kxx,Freese:2026sga,
Wu:2026nhi,Yin:2026jnn,Du:2026guj}, more general inelastic
DM~\cite{Su:2026rwz,DiMauro:2026ldr}, electroweak
multiplets~\cite{Nomura:2026qyq,Visinelli:2026kgt,Smirnov:2026aqk},
and inelastic dark-photon DM~\cite{Yamashita:2026ump}.
Because these scenarios lie close to the inelastic kinematic threshold,
their predictions are highly sensitive to the extreme high-speed tail
of the Galactic DM distribution, including possible enhancements
associated with the Large Magellanic Cloud~\cite{Fan:2026kxx}.
The thermal Higgsino interpretation also faces complementary
constraints: solar capture and the IceCube non-observation of
high-energy solar neutrinos have been argued to require
$\delta>566$~keV~\cite{Pospelov:2026ewn}, while the absence of events
in the higher-energy LZ sideband may provide an additional
constraint~\cite{Rodd:2026tyn}.
Other interpretations based on DM absorption~\cite{Lou:2026idn},
singlino DM~\cite{Chattopadhyay:2026ryw}, elastic axion-portal
scattering~\cite{Unwin:2026rdp}, and atmospheric-neutrino
up-scattering~\cite{Jeesun:2026vzo} have also been proposed.
These developments motivate exploring qualitatively different
mechanisms for generating an isolated high-energy recoil.

In this work, we focus on the elastic MDD interaction and 
explore the impact of a non-standard incident 
DM velocity distribution.  
The strong momentum dependence of the MDD interaction,
$d\sigma/dE_R\propto E_R^2$ up to nuclear-response and kinematic
factors, naturally enhances the relative rate of high-energy recoils. 
Nevertheless, when combined with the standard halo model (SHM)
velocity distribution, the predicted spectrum still contains a
sizable lower-energy component.  
For example, after normalizing the SHM prediction for the 1~TeV DM mass 
to one event in the recoil-energy interval $248\pm23$~keV, 
approximately two events are expected in the $100$--$200$~keV region, 
where no excess is observed. 
This tension motivates a departure from the simple 
SHM velocity distribution.

We show that a harder and narrower DM velocity distribution can
substantially suppress the relative low-energy recoil yield while
maintaining an ${\cal O}(1)$ event near $248$~keV.
We first demonstrate this model-independently using a monochromatic
velocity distribution, which yields a more favorable ratio of high- to
low-energy recoils than the SHM. 
We further show that such a velocity distribution 
allows the observed recoil spectrum to be explained 
with substantially lighter DM. 
As a concrete realization, we consider 
the on-shell mediator scenario of Ref.~\cite{Du:2020ybt}, 
in which halo DM annihilates into on-shell mediators 
that subsequently decay into a new dark-sector species. 
For an appropriate mass hierarchy, 
the final-state particles acquire  
a narrow box-shaped velocity distribution that 
can successfully fit the LZ nuclear recoil spectrum. 
Our results demonstrate that the incident DM velocity distribution can
play an essential role in interpreting high-energy nuclear-recoil
events and that a boosted dark-sector population provides a 
qualitatively different explanation of the LZ event from the
standard-halo inelastic scenarios.

%%%%%%%%%%%%%%%%%%%%%%%%%%%%%%%%%%%%%%
\section{The magnetic dipole-dipole operator}
%%%%%%%%%%%%%%%%%%%%%%%%%%%%%%%%%%%%%%
LZ interpreted its event of interest within a general set of covariant DM--nucleon interactions~\cite{Anand:2013yka,LZ:2024vge},
among which the magnetic dipole-dipole operator can account for the event at $3.4\sigma$ and is given by
\begin{equation}
\cL_{\rm MDD} = d \,
\Big[\bar\psi\, i\sigma^{\mu\nu} \frac{q_\nu}{m_N}\, \psi\Big]
\Big[\bar N\, i\sigma_{\mu\alpha} \frac{q^\alpha}{m_N}\, N\Big] \,,
\label{eq:MDD}
\end{equation}
where $N$ denotes a nucleon, 
$m_N$ is the nucleon
mass,
$q$ is the momentum transfer, 
and $d$ is the corresponding Wilson coefficient of mass dimension $-2$.
We assume an isoscalar interaction throughout so that the proton and neutron couplings are identical.

In the nonrelativistic effective field theory (NREFT), Eq.~\eqref{eq:MDD}
can be expressed in terms of the standard operators~\cite{Anand:2013yka}
\begin{equation}
\cL_{\rm MDD} = 
4\,d \Big[\frac{q^2}{m_N^2}\,\mathcal O_4 - \mathcal O_6\Big]\,,
\label{eq:MDDnr}
\end{equation}
where $\mathcal O_4 = \vec S_\psi\cdot\vec S_N$ and
$\mathcal O_6 = (\vec S_\psi\cdot\vec q/m_N)(\vec S_N\cdot\vec q/m_N)$, with $\vec S_\psi$ ($\vec S_N$) denoting the DM (nucleon) spin.

For DM with the velocity $v$ scattering from a nuclear isotope target $T$
with mass $m_T$ and spin $J_T$, the differential cross section is given by
~\cite{Anand:2013yka}
\begin{equation}
\frac{\dd\sigma_T}{\dd E_R}
= \frac{32\,m_T^3d^2}{v^2(2J_T+1)m_N^4}
E_R^2\,W_T^{\Sigma'}(q)\,,
\label{eq:dsig}
\end{equation}
where $q^2 = 2 m_T E_R$ and
$W_T^{\Sigma'}(q)$ denotes the isoscalar transverse spin
nuclear response function, taken from the shell-model tables in Ref.~\cite{Gorton:2022eed},
which were computed using the GCN5082 interaction~\cite{Caurier:2007wq,Caurier:2010az}\footnote{
With this nuclear response function, we can reproduce the NR spectrum for
magnetic dipole-dipole interactions in Ref.~\cite{LZ:2026axp}.}.

The explicit $E_R^2$ dependence enhances the scattering rate at larger recoil energies, making the MDD interaction an attractive benchmark for explaining the LZ excess event. 
The MDD interaction is also spin dependent, so only nuclei
with nonzero spin contribute to the scattering. 
Consequently, the odd-mass isotopes $^{129}$Xe and $^{131}$Xe provide the relevant scattering targets in the LZ detector.

The differential recoil rate per unit detector mass for the MDD interaction is
\begin{equation}
\frac{\mathrm{d}\mathcal R}{\mathrm{d}E_R}
= \sum_{T}
N_T \int_{v > v_{\min}(E_R)} \mathrm{d}v\;
\frac{\mathrm{d}\Phi_\psi}{\mathrm{d}v}\,
\frac{\mathrm{d}\sigma_T}{\mathrm{d}E_R}(v, E_R) \,,
\label{eq:rate}
\end{equation}
where $N_T=\xi_T/m_T$ is the number of target nuclei per unit detector mass,
with $\xi_T$ denoting the isotopic mass fraction ($\xi_{{}^{129}\mathrm{Xe}}=0.264$ and
$\xi_{{}^{131}\mathrm{Xe}}=0.212$).
The minimum DM velocity required to produce a recoil energy $E_R$ is
$v_{\min}(E_R)=\sqrt{m_T E_R/(2\mu_{\psi T}^2)}$ with  $\mu_{\psi T}=m_\psi m_T/(m_\psi+m_T)$ being the DM--nucleus reduced mass.
The differential DM flux is
$\mathrm{d}\Phi_\psi/\mathrm{d}v=\Phi_\psi f(v)$,
where $\Phi_\psi$ is the total incident DM flux and $f(v)$ is the normalized
DM velocity distribution, satisfying $\int \mathrm{d}v\,f(v)=1$.
For the standard halo model,
the DM velocity distribution is described
by a truncated Maxwellian distribution ~\cite{Lewin:1995rx}.

Integrating the differential recoil rate over the recoil energy gives the expected
number of signal events,
\begin{equation}
N_s = \mathcal{E} \int \mathrm{d}E_R\,
\frac{\mathrm{d}\mathcal R}{\mathrm{d}E_R}\,\epsilon(E_R)\,,
\label{eq:Ns}
\end{equation}
where $\epsilon(E_R)$ is the LZ detection efficiency and
$\mathcal{E}=2.84~\mathrm{t}\!\cdot\!\mathrm{yr}$ is the exposure
~\cite{LZ:2026axp}.

%%%%%%%%%%%%%%%%%%%%%%%%%%%%%%%%%%%%%%
\section{Model-Independent Monochromatic-Velocity Analysis}
%%%%%%%%%%%%%%%%%%%%%%%%%%%%%%%%%%%%%%
Although the MDD interaction with the SHM velocity distribution can provide a
sizable contribution to the observed high-energy event, it also predicts a
significant number of events at lower recoil energies, where no excess is
observed. 
This suggests that DM particles with higher velocities may be favored.

As a model-independent benchmark, we first consider a monochromatic
($\delta$-function) velocity distribution,
$f(v) = \delta(v-v_*)$, where $v_*$ denotes the boosted DM velocity. 
The differential rate in \cref{eq:rate} then becomes
\begin{equation}
\frac{d\mathcal{R}}{dE_R} = \Phi_\psi \sum_T N_T\, \frac{d\sigma_T}{dE_R}(v_*, E_R)\, \Theta\!\left(E_R^{\max}(v_*) - E_R\right),
\label{eq:rate_monochromatic}
\end{equation}
where $E_R^{\max} = 2\mu_{\psi T}^2 v_*^2/m_T$ is the kinematic endpoint at fixed $v_*$.

\begin{figure}[t]
\centering
\figfile{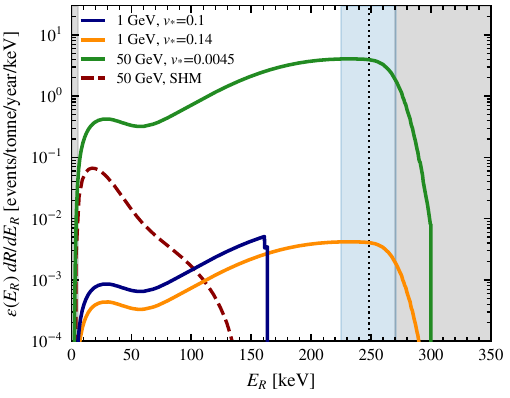}
\caption{Nuclear recoil energy spectra for DM with magnetic dipole-dipole interactions, assuming a monochromatic velocity distribution (solid curves) and the SHM (dashed curve). The solid navy, orange, and green curves correspond to $(m_\psi,v_*)=(1~\mathrm{GeV},0.1)$, $(1~\mathrm{GeV},0.14)$, and $(50~\mathrm{GeV},0.0045)$, respectively, while the dashed red curve shows the SHM prediction for $m_\psi=50~\mathrm{GeV}$. We fix $d=1/m_v^2$ with $m_v = 246.2$ GeV and $\Phi_\psi=10^7~\mathrm{cm}^{-2} \, \mathrm{s}^{-1}$. The light blue shaded region indicates the observed recoil-energy range of the LZ excess event.
}
\label{fig:spectra}
\end{figure}

Figure~\ref{fig:spectra} shows the LZ detection-efficiency-weighted nuclear recoil spectra for monochromatic DM for different DM masses $m_\psi$ and velocities $v_*$. 
We find that, apart from determining the kinematic endpoint, neither $m_\psi$ nor $v_*$ modifies the shape of the normalized recoil spectrum.
Compared with the SHM prediction, a sufficiently large $v_*$ can
extend the recoil spectrum into the region of LZ excess event, even
for relatively light DM masses, whereas the corresponding SHM
spectrum is unable to populate this high-energy region.

Requiring the kinematic endpoint to exceed the observed recoil energy,
$E_R^{\max}\gtrsim248~\mathrm{keV}$, gives
\begin{equation}
E_R^{\max}
=
\frac{2\mu_{\psi T}^2v_*^2}{m_T}
\gtrsim 248~\mathrm{keV},
\end{equation}
or equivalently
$\mu_{\psi T} v_* \gtrsim 0.12~\mathrm{GeV}$ for xenon targets.

To quantify how well a model fits the LZ excess, we define the signal bin as $[225,271]~\text{keV}$ ($\pm1\sigma$ around $248\pm23~\text{keV}$) and the control bin as $[100,200]~\text{keV}$, which contains no background events~\cite{LZ:2026axp} and lies more than $1\sigma$ below the signal region.
The observed counts are $n_{\rm sig}=1$ in the signal bin and $n_{\rm ctl}=0$ in the control bin, with negligible background.

For each parameter point, the expected counts in the control and signal bins, denoted by $N_{\rm ctl}$ and $N_{\rm sig}$, respectively, enter the two-bin Poisson likelihood as
\begin{equation}
    \mathcal{L} = \prod_{i=\rm sig,ctl} \mathrm{Poisson}(n_i|N_i).
\end{equation}
Profiling the signal normalization to match the single event fixes $N_{\rm sig}$, leaving the control-bin constraint. With zero observed events, the likelihood ratio gives
\begin{equation}
    \Delta\chi^2 = 2N_{\rm ctl} \le 4.61 \quad (90\%~\mathrm{CL}),
    \label{eq:chi2}
\end{equation}
i.e., $N_{\rm ctl}\le2.305$. 

\begin{figure}[ht]
\centering
\includegraphics[width=0.9\linewidth]{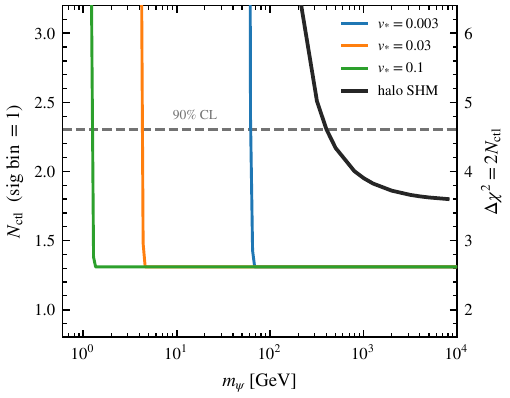}    
\caption{
Expected number of events in the control bin as a function of the dark matter mass. The expected number of events in the signal bin is fixed to one.
The navy, orange and green curves represent the monochromatic boosts with $v_* = 0.003$, $0.03$ and $0.1$, respectively, while the black curve shows the SHM case.
The dashed horizontal line marks the 90\%~CL boundary $N_{\rm ctl} = 2.305$.  
}
\label{fig:Nctl}
\end{figure}

The expected number of events and $\Delta\chi^2$ as functions of $m_\psi$ are shown in \cref{fig:Nctl}, with the flux and cross section profiled at each $m_\psi$ to obtain the best fit.
The boost substantially improves the fit
and extends the allowed mass range to lower masses.
For the SHM, $N_{\rm ctl}$ decreases from
$3.33$ at $m_\psi=200~\rm GeV$ to an asymptotic value of $1.78$, crossing the 90\%~CL boundary at $m_\psi\simeq400~\rm GeV$. 
The monochromatic velocity distribution instead gives $N_{\rm ctl}=1.31$ at sufficiently large $m_\psi$, corresponding to $\Delta\chi^2=2.6$, below both the 90\%~CL boundary.
More importantly, the boost extends the allowed mass range to much lower masses, with the minimum masses reaching $m_\psi\simeq 60$, $4.2$, and $1.2~\rm GeV$ for $v_*=0.003$, $0.03$, and $0.1$, respectively, at which $E_R^{\rm max}$ coincides with the recoil energy of the LZ excess event.

%%%%%%%%%%%%%%%%%%%%%%%%%%%%%%%%%%%%%%
\section{On-shell mediator realization}
%%%%%%%%%%%%%%%%%%%%%%%%%%%%%%%%%%%%%%
A concrete realization of the required narrow boost is provided by the
on-shell mediator model (OMM) of Ref.~\cite{Du:2020ybt}, in which halo DM annihilates via $\chi\chi\to\phi\phi$, followed by
$\phi\to\psi\psi$ 
producing boosted $\psi$ particles. 
The boosted $\psi$ subsequently scatters off nuclei through the MDD interaction as shown in~\cref{eq:MDD}. 
Here we assume the mass hierarchy
$m_\chi>m_\phi>2m_\psi$. Both $\chi$ and $\psi$ are stable and in principle can contribute to the total DM relic abundance. In this analysis, we assume that the relic abundance is dominated by $\chi$, with the contribution from $\psi$ being negligible.

Isotropic two-body decays of the on-shell
$\phi$ generate a box-shaped $\psi$ energy spectrum,
$E_-<E_\psi<E_+$, with
\begin{equation}
E_\pm=\frac{m_\chi}{2}(1\pm xy),
\end{equation}
where $x=\sqrt{1-{m_\phi^2}/{m_\chi^2}}$ and $
y=\sqrt{1-{4m_\psi^2}/{m_\phi^2}}$.
The corresponding normalized velocity distribution is
\begin{equation}
f(v)=
\frac{\sqrt{(1-x^2)(1-y^2)}}{2xy}
\frac{v}{(1-v^2)^{3/2}},
\end{equation}
for $v_-\leq v\leq v_+$, 
where
$v_\pm=\sqrt{1-m_\psi^2/E_\pm^2}$.
It is convenient to further parameterize the endpoints as
$v_\pm=\bar v(1\pm\epsilon)$
where
$\bar v=\frac{x(1-y^2)}{1-x^2y^2}$
denotes the center of the box-shaped velocity distribution, and
$\epsilon=\frac{y(1-x^2)}{x(1-y^2)}$
is its relative half-width.

\begin{figure}[ht]
\centering
\figfile{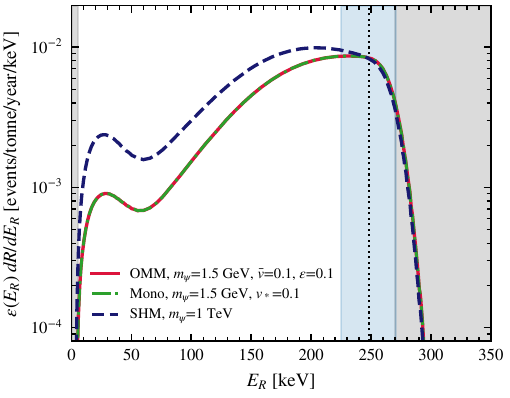}
\caption{
Nuclear recoil energy spectra, normalized to one event in the signal bin
$[225,271]$~keV (blue band). The red solid curve shows the on-shell mediator
model benchmark point $(m_\psi=1.5~\mathrm{GeV},\,\bar v=0.1,\,\epsilon=0.1)$,
while the green dash-dotted curve shows a monochromatic boost with
$v_*=0.1$ and $m_\psi=1.5$~GeV. The navy dashed curve shows the
SHM spectrum at $m_\psi=1$~TeV.
}
\label{fig:benchmark}
\end{figure}

We give a benchmark point in the on-shell mediator model (OMM) with
$m_\psi=1.5~\mathrm{GeV}$, $\bar v=0.1$, and $\epsilon=0.1$. 
Figure~\ref{fig:benchmark} shows the corresponding recoil spectrum, with the overall normalization chosen such that the expected number of events in the signal region is 1. 
The resulting spectrum closely resembles that of the monochromatic scenario with $m_\psi=1.5~\mathrm{GeV}$ and $v_*=0.1$. 
As shown in \cref{fig:benchmark}, the benchmark generates the high-energy recoil feature associated with the LZ excess event while substantially suppressing the low-energy recoil spectrum relative to the SHM prediction.

Next, we show the flux of boosted $\psi$ particles produced 
from $\chi\chi$ annihilations in the Galactic Center, following Ref.~\cite{Du:2020ybt}. 
We assume an NFW profile for the Milky Way DM halo,
\begin{equation}
\rho_\chi(r)=\rho_{s} \frac{\left(r / r_{s}\right)^{-\gamma}}{\left(1+r / r_{s}\right)^{3-\gamma}},
\end{equation}
where we take $\gamma=1$, $\rho_{s}=0.31$~GeV/cm$^3$, and $r_{s}=21$~kpc. 
The resulting flux of $\psi$ is
\begin{equation}
\Phi_\psi = \frac{\langle\sigma v\rangle}{2\pi m_\chi^2}\,J\,,
\label{eq:flux}
\end{equation}
where the total $J$-factor is defined as $J=\int d \Omega \int d s \rho_{\chi}^{2} 
\simeq  10^{23}$ $\mathrm{GeV^2/cm^5}$, 
and $\langle\sigma v\rangle$ denotes the $\chi\chi$ annihilation cross section. 
For $m_{\chi}\simeq 3$~GeV and the canonical thermal annihilation cross section, $\langle\sigma v\rangle=3\times10^{-26}$~cm$^{3}$/s, 
the total $\psi$ flux is
$\Phi_\psi \simeq 5.3\times10^{-5}$~cm$^{-2}$s$^{-1}$. 
For this flux, obtaining one event in the signal bin requires
$d \simeq 4.4\times10^5\,m_v^{-2}$.
This corresponds to $1.3$ events in the control bin, which is significantly below the $\sim 2$ events expected for the SHM case.

To assess Earth attenuation, we first express the required coupling as
the total cross section on a free, pointlike nucleon at the benchmark
speed.  Integrating the leading nonrelativistic interaction in
\cref{eq:MDDnr} gives
\begin{equation}
\sigma_{\psi N}(\bar v)
= \frac{32\,d^2\,\mu_{\psi N}^6\,\bar v^4}{3\pi m_N^4}
\simeq 3 \times10^{-31}~\mathrm{cm^2},
\label{eq:sigmaN}
\end{equation}
where $\mu_{\psi N}=m_\psi m_N/(m_\psi+m_N)$, with equal proton and
neutron cross sections in the isoscalar limit.
Previous studies find much weaker elastic stopping for SD than for
SI interactions, at $10^{-24}$~cm$^2$, above the nucleon's geometric
area~\cite{QUESTDMC:2025attenuation, Bringmann:2018cvk}.
Also, LZ's boosted DM study~\cite{LZ:2025CRDM} and the Monte Carlo framework it built on~\cite{Xia:2021vbz} 
illustrate that the attenuation effect mostly affect highly energetic part of the boosted flux, 
while our narrow box flux has $T_\psi\ll m_\psi$ and therefore lies outside
this ultrarelativistic regime.

%%%%%%%%%%%%%%%%%%%%%%%%%%%%%%%%%%%%%%
\section{Conclusion}
%%%%%%%%%%%%%%%%%%%%%%%%%%%%%%%%%%%%%%
The recoil spectrum from magnetic dipole-dipole interactions between dark matter and nucleons is strongly enhanced at large momentum transfer, making this interaction particularly well suited for generating a localized high-energy feature and providing a viable interpretation of the high-energy nuclear-recoil excess recently observed by LZ. However, when the dark matter follows the standard halo velocity distribution, the resulting recoil spectrum also predicts an sizable number of lower-energy events.

In this Letter, we show that a boosted velocity distribution of dark matter shifts the recoil spectrum toward the observed high-energy region while suppressing the associated low-energy contribution, thereby providing a better fit to the LZ excess. 
Using a model-independent monochromatic velocity distribution, we find that a broad range of dark matter masses  and boost velocities can reproduce an event observed at LZ  without overproducing low-energy events.
We further provide a concrete realization for such a boosted through the cascade
$\chi\chi\to\phi\phi\to4\psi$ in the on-shell mediator model. In particular, the boosted $\psi$ particles acquire a narrow, box-shaped velocity distribution. For a benchmark point with $m_\psi = 1.5$ GeV, the center velocity of the box-shaped $\bar{v} = 0.1$ and its relative haft-width $\epsilon = 0.1$, the recoil spectrum closely reproduces the results of the same mass dark matter with monochromatic velocity $v_* = 0.1$. 
We also shown that for this benchmark point, the required cross section between $\psi$ and nucleon is sufficiently small that attenuation in the Earth is not expected to qualitatively modify the signal.

Our results illustrate the important role that the dark matter velocity distribution can play in interpreting nuclear-recoil anomalies, particularly when the observed events lie close to the kinematic edge of conventional halo dark matter. More broadly, boosted dark sector populations provide a well-motivated possibility for generating high-recoil signals that are difficult to accommodate under the standard halo assumption. It would be interesting to extend this analysis to a broader basis of dark matter-nucleon effective operators and to investigate the associated astrophysical, cosmological, and experimental constraints.

%%%%%%%%%%%%%%%%%%%%%%%%%%%%%%%%%%%%%%
\section{Acknowledgments}
%%%%%%%%%%%%%%%%%%%%%%%%%%%%%%%%%%%%%%
The work is supported in part by the 
National Natural Science Foundation of China under Grant 
No.\ 12275128.
It is also partially supported by the National Science and Technology Council, the Ministry of Education (Higher Education Sprout Project NTU-114L104022-1), the National Center for Theoretical Sciences of Taiwan, and the Vietnam National Foundation for Science and Technology Development (NAFOSTED) under Grant No.~103.01-2023.50 (VQT).

\bibliography{ref}

@article{LZ:2026axp,
    author        = "Akerib, D. S. and others",
    collaboration = "LZ",
    title         = "{Search for dark matter particle interactions in an extended nuclear recoil energy window with the LUX-ZEPLIN (LZ) experiment}",
    eprint        = "2609.02823",
    archivePrefix = "arXiv",
    primaryClass  = "hep-ex",
    month         = "9",
    journal       = "",
    year          = "2026"
}

@misc{LZtalk2026,
    author        = "{LZ Collaboration}",
    title         = "{Dark matter EFT nuclear recoil search at higher energies}",
    howpublished  = "contribution to TeVPA 2026",
    year          = "2026"
}

@article{Anand:2013yka,
    author        = "Anand, Nikhil and Fitzpatrick, A. Liam and Haxton, W. C.",
    title         = "{Weakly interacting massive particle-nucleus elastic scattering response}",
    eprint        = "1308.6288",
    archivePrefix = "arXiv",
    primaryClass  = "hep-ph",
    doi           = "10.1103/PhysRevC.89.065501",
    journal       = "Phys. Rev. C",
    volume        = "89",
    number        = "6",
    pages         = "065501",
    year          = "2014"
}

@article{LZ:2024vge,
    author        = "Aalbers, J. and others",
    collaboration = "LZ",
    title         = "{Constraints on Covariant Dark-Matter{\textendash}Nucleon Effective Field Theory Interactions from the First Science Run of the LUX-ZEPLIN Experiment}",
    eprint        = "2404.17666",
    archivePrefix = "arXiv",
    primaryClass  = "hep-ex",
    reportNumber  = "FERMILAB-PUB-24-0760-V",
    doi           = "10.1103/PhysRevLett.133.221801",
    journal       = "Phys. Rev. Lett.",
    volume        = "133",
    number        = "22",
    pages         = "221801",
    year          = "2024"
}

@article{Du:2020ybt,
    author        = "Du, Mingxuan and Liang, Jinhan and Liu, Zuowei and Tran, Van Que and Xue, Yilun",
    title         = "{On-shell mediator dark matter models and the Xenon1T excess}",
    eprint        = "2006.11949",
    archivePrefix = "arXiv",
    primaryClass  = "hep-ph",
    doi           = "10.1088/1674-1137/abc244",
    journal       = "Chin. Phys. C",
    volume        = "45",
    number        = "1",
    pages         = "013114",
    year          = "2021"
}

@article{Bringmann:2018cvk,
    author        = "Bringmann, Torsten and Pospelov, Maxim",
    title         = "{Novel direct detection constraints on light dark matter}",
    eprint        = "1810.10543",
    archivePrefix = "arXiv",
    primaryClass  = "hep-ph",
    doi           = "10.1103/PhysRevLett.122.171801",
    journal       = "Phys. Rev. Lett.",
    volume        = "122",
    number        = "17",
    pages         = "171801",
    year          = "2019"
}

@article{Gorton:2022eed,
    author        = "Gorton, Oliver C. and Johnson, Calvin W. and Jiao, Changfeng and Nikoleyczik, Jonathan",
    title         = "{dmscatter: A fast program for WIMP-nucleus scattering}",
    eprint        = "2209.09187",
    archivePrefix = "arXiv",
    primaryClass  = "nucl-th",
    doi           = "10.1016/j.cpc.2022.108597",
    journal       = "Comput. Phys. Commun.",
    volume        = "284",
    pages         = "108597",
    year          = "2023"
}

@article{Su:2026rwz,
    author        = "Su, Liangliang and Yang, Jin Min and Yang, Wen-Na",
    title         = "{Inelastic Dark Matter Signature at High Recoil Energy in LUX-ZEPLIN and CRESST}",
    eprint        = "2609.01475",
    archivePrefix = "arXiv",
    primaryClass  = "hep-ph",
    month         = "9",
    journal       = "",
    year          = "2026"
}

@article{Fan:2026kxx,
    author        = "Fan, JiJi and Reece, Matthew",
    title         = "{Higgsino Above the Sea of Fog}",
    eprint        = "2609.01504",
    archivePrefix = "arXiv",
    primaryClass  = "hep-ph",
    month         = "9",
    journal       = "",
    year          = "2026"
}

@article{Freese:2026sga,
    author        = "Freese, Katherine and Theodosopoulos, Dionysios P.",
    title         = "{Higgsino Dark Matter Interpretation of the LUX-ZEPLIN 248 keV Nuclear-Recoil Event}",
    eprint        = "2609.01583",
    archivePrefix = "arXiv",
    primaryClass  = "hep-ph",
    month         = "9",
    journal       = "",
    year          = "2026"
}

@article{Wu:2026nhi,
    author        = "Wu, Lei and Zhang, Yang and Zhu, Bin",
    title         = "{TeV Higgsino Dark Matter from LZ Nuclear Recoil to Fermi-LAT Gamma Rays}",
    eprint        = "2609.01590",
    archivePrefix = "arXiv",
    primaryClass  = "hep-ph",
    month         = "9",
    journal       = "",
    year          = "2026"
}

@article{Lou:2026idn,
    author        = "Lou, Yuanchao and Lu, Chih-Ting",
    title         = "{Fermionic Dark Matter Absorption and the High-Energy Event in LUX-ZEPLIN}",
    eprint        = "2609.01592",
    archivePrefix = "arXiv",
    primaryClass  = "hep-ph",
    month         = "9",
    journal       = "",
    year          = "2026"
}

@article{Yin:2026jnn,
    author        = "Yin, Wen",
    title         = "{A PQ-Symmetric High-Scale SUSY Interpretation of the LZ High-Energy Recoil}",
    eprint        = "2609.01892",
    archivePrefix = "arXiv",
    primaryClass  = "hep-ph",
    month         = "9",
    journal       = "",
    year          = "2026"
}

@article{Nomura:2026qyq,
    author        = "Nomura, Yasunori",
    title         = "{Dark Matter as the Z{\_}2 Partner of the Standard Model Higgs Boson}",
    eprint        = "2609.02505",
    archivePrefix = "arXiv",
    primaryClass  = "hep-ph",
    reportNumber  = "RIKEN-iTHEMS-Report-26",
    month         = "9",
    journal       = "",
    year          = "2026"
}

@article{DiMauro:2026ldr,
    author        = "Di Mauro, Mattia",
    title         = "{Dark Matter at the Kinematic Edge: Interpreting the 248 keV LZ Nuclear-Recoil Candidate}",
    eprint        = "2609.02608",
    archivePrefix = "arXiv",
    primaryClass  = "hep-ph",
    month         = "9",
    journal       = "",
    year          = "2026"
}

@article{Pospelov:2026ewn,
    author        = "Pospelov, Maxim and Ramani, Harikrishnan",
    title         = "{Strong Constraints on Higgsino Dark Matter from Solar Capture}",
    eprint        = "2609.02775",
    archivePrefix = "arXiv",
    primaryClass  = "hep-ph",
    month         = "9",
    journal       = "",
    year          = "2026"
}

@article{Visinelli:2026kgt,
    author        = "Visinelli, Luca",
    title         = "{A Peccei--Quinn Origin for Inelastic Electroweak Dark Matter after LUX-ZEPLIN}",
    eprint        = "2609.02807",
    archivePrefix = "arXiv",
    primaryClass  = "hep-ph",
    month         = "9",
    journal       = "",
    year          = "2026"
}

@article{Yamashita:2026ump,
    author        = "Yamashita, Kimiko",
    title         = "{Inelastic Dark Photon Dark Matter for the LUX-ZEPLIN High-Recoil Event and the Galactic Halo Gamma-Ray Excess}",
    eprint        = "2609.02868",
    archivePrefix = "arXiv",
    primaryClass  = "hep-ph",
    month         = "9",
    journal       = "",
    year          = "2026"
}

@article{Unwin:2026rdp,
    author        = "Unwin, James",
    title         = "{Axion Portal Dark Matter and the LUX-ZEPLIN High-Recoil Event}",
    eprint        = "2609.04186",
    archivePrefix = "arXiv",
    primaryClass  = "hep-ph",
    month         = "9",
    journal       = "",
    year          = "2026"
}

@article{Chattopadhyay:2026ryw,
    author = "Chattopadhyay, Utpal and Das, Debottam and Puri, Rahul and Roy, Joydeep",
    title = "{Sub-TeV Singlino Dark Matter in light from Sagittarius A$^\ast$ and LUX-ZEPLIN Nuclear-Recoil Event}",
    eprint = "2609.02994",
    archivePrefix = "arXiv",
    primaryClass = "hep-ph",
    month = "9",
    journal = "",
    year = "2026"
}

@article{Smirnov:2026aqk,
    author = "Smirnov, Juri and Griffith, Spencer and Beacom, John F.",
    title = "{Inelastic Signatures of Electroweak Dark Matter}",
    eprint = "2609.04144",
    archivePrefix = "arXiv",
    primaryClass = "hep-ph",
    month = "9",
    journal = "",
    year = "2026"
}

@article{Du:2026guj,
    author = "Du, Xiaokang and Wang, Fei",
    title = "{TeV Higgsino Interpretation of the LZ High-Recoil Event with Intermediate-Scale Electroweak Gauginos}",
    eprint = "2609.04163",
    archivePrefix = "arXiv",
    primaryClass = "hep-ph",
    month = "9",
    journal = "",
    year = "2026"
}

@article{Rodd:2026tyn,
    author = "Rodd, Nicholas L. and Safdi, Benjamin R. and Slatyer, Tracy R. and Xu, Weishuang Linda",
    title = "{Confronting the Higgsino Interpretation of the LZ Event with the High-Energy Sideband}",
    eprint = "2609.04175",
    archivePrefix = "arXiv",
    primaryClass = "hep-ph",
    month = "9",
    journal = "",
    year = "2026"
}

@article{Jeesun:2026vzo,
    author = "Jeesun, Sk and Majumdar, Anirban",
    title = "{Atmospheric neutrino up-scattering explanation of LZ 2026 excess}",
    eprint = "2609.04185",
    archivePrefix = "arXiv",
    primaryClass = "hep-ph",
    month = "9",
    journal = "",
    year = "2026"
}

@article{Caurier:2007wq,
    author = "Caurier, E. and Menendez, J. and Nowacki, F. and Poves, A.",
    title = "{The Influence of pairing on the nuclear matrix elements of the neutrinoless beta beta decays}",
    eprint = "0709.2137",
    archivePrefix = "arXiv",
    primaryClass = "nucl-th",
    doi = "10.1103/PhysRevLett.100.052503",
    journal = "Phys. Rev. Lett.",
    volume = "100",
    pages = "052503",
    year = "2008"
}

@article{Caurier:2010az,
    author = "Caurier, E. and Nowacki, F. and Poves, A. and Sieja, K.",
    title = "{Collectivity in the light Xenon isotopes: A shell model study}",
    eprint = "1009.3813",
    archivePrefix = "arXiv",
    primaryClass = "nucl-th",
    doi = "10.1103/PhysRevC.82.064304",
    journal = "Phys. Rev. C",
    volume = "82",
    pages = "064304",
    year = "2010"
}

@article{Xia:2021vbz,
    author = "Xia, Chen and Xu, Yan-Hao and Zhou, Yu-Feng",
    title = "{Production and attenuation of cosmic-ray boosted dark matter}",
    eprint = "2111.05559",
    archivePrefix = "arXiv",
    primaryClass = "hep-ph",
    doi = "10.1088/1475-7516/2022/02/028",
    journal = "JCAP",
    volume = "2022",
    number = "02",
    pages = "028",
    year = "2022"
}

@article{LZ:2025CRDM,
    author = "Aalbers, J. and others",
    collaboration = "LZ",
    title = "{New Constraints on Cosmic Ray-Boosted Dark Matter from the LUX-ZEPLIN Experiment}",
    eprint = "2503.18158",
    archivePrefix = "arXiv",
    primaryClass = "hep-ex",
    doi = "10.1103/nr92-jvt3",
    journal = "Phys. Rev. Lett.",
    volume = "134",
    number = "24",
    pages = "241801",
    year = "2025"
}

@article{QUESTDMC:2025attenuation,
    author = "Darvishi, N. and others",
    collaboration = "QUEST-DMC",
    title = "{Dark matter attenuation effects: sensitivity ceilings for spin-dependent and spin-independent interactions}",
    eprint = "2502.10251",
    archivePrefix = "arXiv",
    primaryClass = "hep-ph",
    doi = "10.1088/1475-7516/2025/04/017",
    journal = "JCAP",
    volume = "2025",
    number = "04",
    pages = "017",
    year = "2025"
}

@article{Lewin:1995rx,
    author = "Lewin, J. D. and Smith, P. F.",
    title = "{Review of mathematics, numerical factors, and corrections for dark matter experiments based on elastic nuclear recoil}",
    reportNumber = "RAL-TR-95-024",
    doi = "10.1016/S0927-6505(96)00047-3",
    journal = "Astropart. Phys.",
    volume = "6",
    pages = "87--112",
    year = "1996"
}

\end{document}